\documentclass[prb,aps,twocolumn,showpacs]{revtex4}

\usepackage{epsfig}

\begin{document}

\title{c-axis longitudinal magnetoresistance of the electron-doped superconductor Pr$_{1.85}$Ce$_{0.15}$CuO$_{4}$}

\author{W. Yu}
  \email{weiqiang@umd.edu}
\author{B. Liang}
\author{R. L. Greene}
\affiliation{Center for Superconductivity Research, Department of Physics, University of Maryland, College Park, MD 20742}

\date{\today}

\begin{abstract}

We report c-axis resistivity and longitudinal magnetoresistance measurements of superconducting Pr$_{1.85}$Ce$_{0.15}$CuO$_{4}$ single crystals. In %%@
the temperature range $13K\le T \le 32K$, a negative magnetoresistance is observed at fields just above $H_{C2}$. Our studies suggest that this %%@
negative magnetoresistance is caused by superconducting fluctuations. At lower temperatures ($T\le 13K$), a different magnetoresistance behavior and a %%@
resistivity upturn are observed, whose origin is still unknown. 
 
\end{abstract}

\maketitle

Both electron-doped (n-type) and hole-doped (p-type) cuprate superconductors show many interesting properties. One important question is the %%@
particle-hole symmetry, that is, whether the phenomena observed in the hole-doped cuprates are the same as those observed in the hole-doped cuprates. %%@
The study of this issue may improve our understanding of their normal state properties and the origin of high-temperature superconductivity. 

C-axis transport has been shown to be a useful measurement for electronic properties, and a comparison of c-axis transport in both types of cuprate %%@
superconductors is in progress. For example, a four-fold oscillation of the c-axis angular magnetoresistance has been reported in underdoped n-type %%@
cuprates \cite{Lavrov_prl_92_227003, Fournier_prb_69_220501, Li_prb_71_054505}, which suggests a stripe-like structure as in the p-type cuprates. %%@
C-axis transport is also a good probe of the electron density of states (DOS) at the Fermi surface close to ($\pi$, 0) in the tetragonal cuprates %%@
\cite{Ioffe_prb_58_11631, Su_prb_73_134510}, because of its intrinsic interlayer tunneling nature and a transfer integral effect. A pseudogap, %%@
indicated by a loss of electronic DOS, has been observed by c-axis transport in both the p-type \cite{Szabo_prl_86_5990, Kim_prb_70_144510, %%@
Lavrov_epl_57_267, Ong_prb_1995_751, Shibauchi_prl_86_5763, Ono_prb_70_224521, Vedeneev_prb_62_5997, Vedeneev_prb_70_184524} and the n-type cuprates %%@
\cite{Onose_prb_2004_024504}. However, from the evidence to date it is not clear if the pseudogaps in the two systems are of the same origin. 

In the underdoped (x$<$0.15) n-type cuprates, a rapid decrease of the low-temperature c-axis resistivity, compared to the ab-plane resistivity %%@
\cite{Onose_prb_2004_024504}, is consistent with a coherent transport at ($\pi$, 0) and a high-energy ($\sim$100 meV) pseudogap at ($\pi/2$, $\pi/2$), %%@
as shown by ARPES \cite{Armitage_prl_88_257001}, optical \cite{Onose_prb_2004_024504, Zimmers_EPL_70_225}, and other transport %%@
\cite{Dagan_prl_92_167001} measurements. This pseudogap is likely caused by antiferromagnetic ordering or a spin density wave (SDW) up to a doping %%@
level x=0.15 \cite{Armitage_prl_88_257001, Zimmers_EPL_70_225, Dagan_prl_92_167001, Millis}. The pseudogap in the p-type compounds, as shown by a %%@
large c-axis resistivity upturn and a negative magnetoresistance, is very prominent. Indeed, two types of pseudogaps are suggested in different %%@
regimes \cite{Lavrov_epl_57_267}. A small-energy gap is consistent with a precursor pairing or superconducting fluctuation scenario  \cite{AL_process, %%@
Szabo_prl_86_5990, Kim_prb_70_144510, Lavrov_epl_57_267}. A large-energy pseudogap opens at ($\pi$, 0), which is unlikely to be correlated with %%@
superconductivity although the origin is still being debated \cite{Ong_prb_1995_751, Shibauchi_prl_86_5763, Ono_prb_70_224521, Vedeneev_prb_62_5997, %%@
Vedeneev_prb_70_184524, Lavrov_epl_57_267}. 

In this work, we studied the low temperature c-axis resistivity and longitudinal magnetoresistance of optimally electron-doped  %%@
Pr$_{1.85}$Ce$_{0.15}$CuO$_{4}$ (PCCO) single crystals. For the temperature range $13K\le T\le 32K$, we found a negative magnetoresistance in a field %%@
range $H<H_P(T)$ (defined later). $H_P(T)$ is higher than $H_{C2}(T)$ and increases as temperature decreases. Our detailed studies suggest that this %%@
high-temperature n-MR is caused by superconducting fluctuations in a quasi-2D system. For lower temperatures $T<13K$, a different magnetoresistance %%@
behavior and a resistivity upturn are observed, the origin of which is still unclear.  

\begin{figure}
\includegraphics[width=8cm, height=7cm]{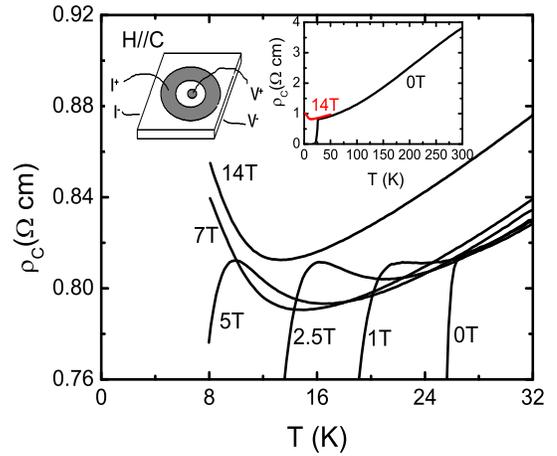}
\caption{\label{rvstb} The temperature dependence of the c-axis resistivity of an optimally doped Pr$_{1.85}$Ce$_{0.15}$CuO$_4$ crystal at various %%@
fields. Inset: The electric contact configuration of the four-wire measurements (left), and the c-axis resistivity at $\mu _0 H=0T$ and $14T$ %%@
respectively (right). }
\end{figure}

The PCCO single crystals were grown by the self-flux method and the usual oxygen reduction procedure was followed to achieve superconductivity %%@
\cite{Balci_prb_70_140508}. The crystals are platelike with size $\sim$0.5$\times$0.5$\times$0.03mm$^3$. The sharp superconducting transition ($T %%@
_{C}=23.5\pm 0.75 K$) found by SQUID magnetization measurements indicates a high crystal quality. For resistivity, a conventional four-wire %%@
measurement was utilized, with two annular contacts painted on either surface of the crystal (see Fig.~\ref{rvstb} inset). Crystals were then mounted %%@
on a rotator and measured in a Quantum Design 14-Tesla PPMS (Physical Property Measurement System).  

The temperature dependence of the c-axis resistivity of an optimally doped PCCO crystal is shown at various longitudinal magnetic fields $H\parallel %%@
C$ in Fig.~\ref{rvstb}. At zero field, the superconducting transition occurs at $T_C\approx 25K$ as indicated by zero resistivity. At low fields ($\mu %%@
_0 H \le 2.5$T), a small upturn feature in the c-axis transport is seen just above the superconducting transition. This resistivity minimum shifts to %%@
lower temperature as the field increases. At high fields ($ H\ge H_{C2}$), the resistivity shows a strong upturn feature as temperature decreases %%@
below $T\approx 13$K. 

\begin{figure}
\includegraphics[width=9cm, height=9cm]{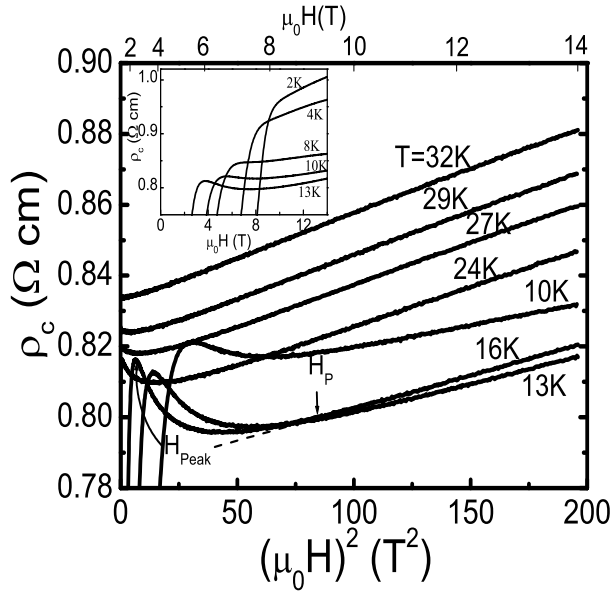}
\caption{\label{rcvsh} The longitudinal magnetoresistance of the c-axis transport of a PCCO crystal with $H\parallel C$ at temperatures $10K\le T \le %%@
32K$. Inset: The low-temperature magnetoresistance at 2K$\le$T$\le$13K.}
\end{figure}

We find that the magnetoresistance behaviors are different in the temperature range above and below $T=13K$. The c-axis longitudinal magnetoresistance %%@
is shown in Fig.~\ref{rcvsh}. For clarity, we plotted $\rho _C (H,T)$ versus $(\mu _0 H)^2$. At T$>$32K, the magnetoresistance, $\Delta \rho _C (H,T) %%@
\equiv \rho _C (H,T)- \rho _C (0,T)$, is positive and proportional to $H^2$ at high temperatures (data not shown). With decreasing temperature from %%@
$32K$ to $13K$, a deviation from the $H^2$ behavior develops at low field, and a negative magnetoresistance (n-MR) appears close to zero field. For %%@
$T<T_C$, the n-MR is more evident, so that a resistivity peak forms as superconductivity is suppressed. The n-MR extends to higher fields as %%@
temperature drops. Below $13K$, a different magnetoresistance behavior is observed, as seen by the suppression of the n-MR and disappearance of the %%@
resistivity peak at T$<$8K  (Fig.~\ref{rcvsh} inset).    

This different magnetoresistive behavior in the temperature range below and above $13K$ can be explained as arising from two contributions. The %%@
high-temperature n-MR comes from superconducting fluctuations, whereas the low-temperature magnetoresistance behavior and the resistivity upturn come %%@
from another mechanism of unknown origin. In the following, we discuss these two contributions separately.

\begin{figure}
\includegraphics[width=8cm, height=7cm]{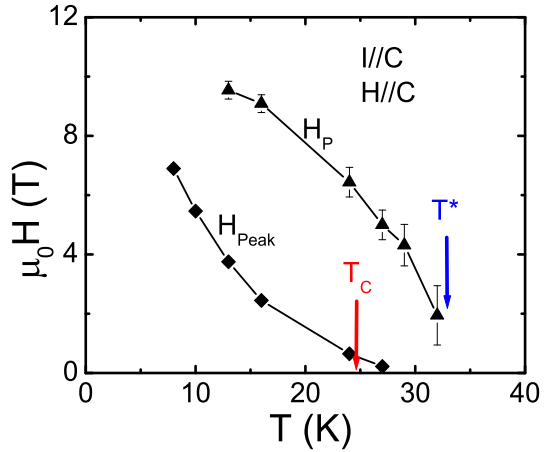}
\caption{\label{hpvsAngle} The mapped temperature dependence of $H_P$ and $H_{Peak}$ as defined in Fig.~\ref{rcvsh}.  }
\end{figure}

We first discuss the origin of the n-MR behavior for T$\ge$13K. In Fig. \ref{rcvsh}, we define $H_{Peak}$ as the field corresponding to the %%@
resistivity peak, and $H_P$ as the lowest field where $\Delta \rho _C (H,T) \sim H^2$ is obeyed. In Fig.~\ref{hpvsAngle}, the values of $H_{Peak}$ and %%@
$H_P$ at different temperatures are shown. Therefore, the n-MR exists between $H_P$ and $H_{peak}$. $H_P$ emerges at $T^*=32K$, which is slightly %%@
above $T_C$. As temperature decreases, the temperature dependence of $H_P$ suggests that it extrapolates to $\mu _0 H_P \sim 10T$ at T=0. We did not %%@
measure $H_{C2}$ on our crystal, but a recent Nernst effect measurement \cite{Wang_science_299_86} on an optimally doped NCCO crystal ($T_C \approx %%@
24.5K$) gave $\mu _0 H_{c2} \approx 6T$ at 13K, which is below our value of $\mu _0 H_P$ $\approx$10T at 13K. The slightly higher value of $T^*$ %%@
($H_P$) than that of $T_C$ ($H_{C2}$), suggests that the c-axis n-MR is associated with superconducting fluctuations. 

\begin{figure}
\includegraphics[width=8cm, height=7cm]{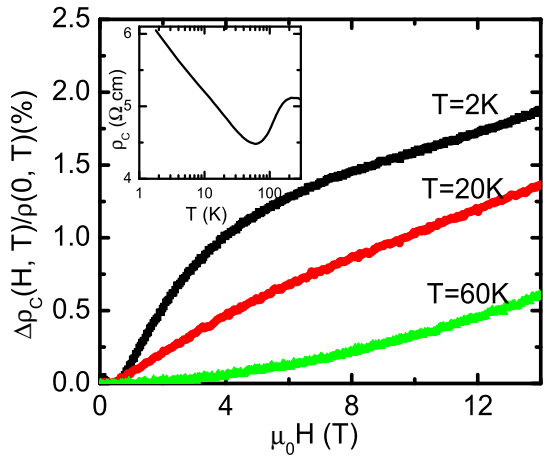}
\caption{\label{rcasgrown} The c-axis longitudinal magnetoresistance of an as-grown, non-superconducting, Pr$_{1.85}$Ce$_{0.15}$CuO$_{4+\delta}$ %%@
crystal. Inset: temperature dependence of the c-axis resistivity.}
\end{figure}

We now show that the doping dependence of the n-MR supports our view that the n-MR for T$\ge$13K originates from superconducting fluctuations. %%@
Although PCCO crystals with other doping levels were not available for our c-axis measurements, we measured an as-grown (unannealed), %%@
non-superconducting, Pr$_{1.85}$Ce$_{0.15}$CuO$_{4+\delta}$ crystal. Neutron scattering studies have shown that this unannealed crystal is equivalent %%@
to an antiferromagnetic, annealed, Pr$_{1.88}$Ce$_{0.12}$CuO$_{4}$ crystal \cite{Mang_prl_93_027002}. As shown in the inset of Fig.~\ref{rcasgrown}, a %%@
resistivity upturn is clearly seen at $T\le 60K$, which is indicative of antiferromagnetic ordering \cite{Onose_prb_2004_024504}. Moreover, a positive %%@
magnetoresistive behavior is seen at all temperatures below $T=60K$ as shown in Fig.~\ref{rcasgrown}. {\it The absence of the n-MR in this effectively %%@
underdoped crystal suggests that the n-MR is unlikely to be correlated with antiferromagnetism or a spin pseudogap state.} The appearance of the n-MR %%@
only in the superconducting crystals strongly supports our interpretation of superconducting fluctuations as the origin of the n-MR. 

\begin{figure}
\includegraphics[width=8cm, height=7cm]{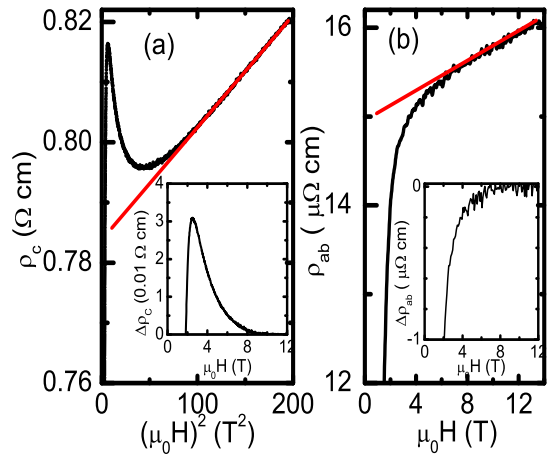}
\caption{\label{rab&c} (a)The c-axis longitudinal magnetoresistance of a PCCO crystal at $T=16K$. The solid line is a fit to high-field  %%@
magnetoresistance ($\propto H^2$). Inset: the low-field n-MR found by subtracting off the high-field fit. (b) The ab-plane transverse %%@
magnetoresistance of the PCCO crystal at $T=16K$. The solid line is a fit to the high-field magnetoresistance ($\propto H$). Inset: the low-field %%@
positive magnetoresistance found by subtracting off the high-field fit. }
\end{figure}

A qualitative comparison between the ab-plane and the c-axis magnetoresistance confirms our interpretation of superconducting fluctuations for %%@
T$\ge$13K. We measured the transverse magnetoresistance of the ab-plane transport with $H\parallel C$ and compared it with the c-axis longitudinal %%@
magnetoresistance at a specific temperature 16K as shown in Fig.~\ref{rab&c}. The ab-plane magnetoresistance is positive and increases linearly with %%@
field above $8T$. To estimate the deviation of the low-field magnetoresistance, we subtracted a fit to the high-field magnetoresistance for both %%@
transports, as shown in Fig.~\ref{rab&c}. Below $\mu _0H=$8T, the c-axis transport shows a negative magnetoresistance (Fig.~\ref{rab&c} (a) inset), %%@
whereas the ab-plane transport shows a positive magnetoresistance (Fig.~\ref{rab&c} (b) inset). This distinctive c-axis and the ab-plane %%@
magnetoresistance can be explained by an Aslamazov-Larkin (AL) contribution from fluctuating  Cooper pairs and a density of states (DOS) contribution %%@
from electrons in a highly anisotropic superconductor \cite{AL_process, AV2}. When increasing the field above $H_{C2}$, the suppression of %%@
superconducting fluctuations causes a decrease of the ab-plane conductivity from the reduced AL contribution, and an increase of the c-axis tunneling %%@
conductivity from the resulting increase of the electronic DOS. 

To summarize, the appearance of a c-axis n-MR only in superconducting PCCO crystals in fields (temperature) just above $H_{C2}$ ($T_C$) suggests that %%@
the n-MR is due to superconducting fluctuations. The contrasting c-axis and ab-plane magnetoresistance in the same field (temperature) range can be %%@
understood by an AL process and a reduction of electronic DOS due to superconducting fluctuations in a quasi-2D system. The c-axis n-MR of PCCO is %%@
rather similar to that found in the p-type superconductor Bi-2201\cite{Lavrov_epl_57_267}, which has been interpreted to be caused by superconducting %%@
fluctuations. The only difference is that superconducting fluctuations in PCCO seem to occur in a smaller range of temperature and field. Our n-MR %%@
does not seem to be related to the low-energy normal-state tunneling gap \cite{Kleefisch_prb_63_100507, Biswas_prb_64_104519, alff_nature, %%@
Dagan_tunneling}, because the tunneling gap is also found for $x=0.11$ where we see only positive magnetoresistance. 

Now we discuss the low-temperature c-axis resistivity and magnetoresistance. Both a different magnetoresistance behavior (see Fig.~\ref{rcvsh} inset) %%@
and a strong resistivity upturn (see Fig.~\ref{rvstb}) occur below 13K. This suggests that the resistivity and the magnetoresistance are correlated.  %%@
It may be related to the same mechanism which causes the resistivity upturn in the ab-plane transport \cite{Dagan_prl_94_057005}. Recently Kawakami %%@
{\it et al.} \cite{Kawakami_prl_2005_017001} reported a negative c-axis magnetoresistance (n-MR) in optimally doped n-type %%@
Sm$_{1.85}$Ce$_{0.15}$CuO$_{4}$ (SCCO) up to 40T and down to 0.5K. Since n-MR has been observed in both the p-type and the n-type cuprates, they %%@
associated the n-MR with a universal Zeeman-splitting effect of a spin pseudogap in both systems. For comparison, the absence of the n-MR and the %%@
resistivity peak below 14T in our PCCO crystals (see Fig.~\ref{rcvsh} inset) is different from that observed by Kawakami {\it et al.} %%@
\cite{Kawakami_prl_2005_017001} in SCCO. Currently, we are not able to verify their spin pseudogap interpretation of the n-MR because we lack %%@
sufficient high-field, low-temperature, data. As presented earlier, our high-temperature n-MR is most likely caused by superconducting fluctuations. 
 
In summary, we have studied the c-axis resistivity and magnetoresistance on optimally doped Pr$_{1.85}$Ce$_{0.15}$CuO$_{4}$ crystals. Different %%@
resistivity and magnetoresistance behavior in two temperature ranges, suggests that two contributions should be considered to understand the c-axis %%@
transport. For 13K$\le$T$\le$32K, a n-MR, which exists only in the superconducting crystals in fields just above $H_{C2}$, is most likely caused by %%@
superconducting fluctuations. A distinctive c-axis negative-magnetoresistance and an ab-plane positive-magnetoresistance is found, which can be %%@
explained by an Aslamazov-Larkin process and a reduction of electronic DOS in a highly anisotropic superconductor. For T$<$13K, a different  %%@
magnetoresistance behavior and a resistivity upturn are observed, the origin of which is unclear at present. 

This work is supported by the NSF under contract DMR-0352735. The authors would like to acknowledge A. J. Millis, V. M. Yakovenko, A. M. Chubukov, and %%@
T. Xiang for beneficial discussions.

%\bibliography{c_transport}

\end{document}